\documentclass{article}

\usepackage[final]{d3s3_neurips_2024}

\usepackage[utf8]{inputenc} 
\usepackage[T1]{fontenc}    
\usepackage{hyperref}       
\usepackage{url}            
\usepackage{booktabs}       
\usepackage{amsfonts}       
\usepackage{nicefrac}       
\usepackage{microtype}      
\usepackage{xcolor}         
\usepackage{amsmath}
\usepackage{bm}
\usepackage{graphicx}
\usepackage{natbib}
\setcitestyle{numbers}

\title{Convolutional Hierarchical Deep Learning Neural Networks-Tensor Decomposition (C-HiDeNN-TD): a scalable surrogate modeling approach for large-scale physical systems}

\author{
 Jiachen Guo \\
 Theoretical and Applied Mechanics\\
 Northwestern University\\
 Evanston, Illinois, 60201 \\
 \texttt{jiachen.guo@northwestern.edu} \\
  \And
  Chanwook Park \\
 Department of Mechanical Engineering\\
 Northwestern University\\
 Evanston, Illinois, 60201 \\
 \texttt{chanwook.park@northwestern.edu} \\
 \And
  Xiaoyu Xie \\
 Department of Mechanical Engineering\\
 Northwestern University\\
 Evanston, Illinois, 60201 \\
 \texttt{xiaoyu.xie@northwestern.edu} \\
\And
 Zhongsheng Sang \\
 Theoretical and Applied Mechanics\\
 Northwestern University\\
 Evanston, Illinois, 60201 \\
 \texttt{zhongsheng.sang@northwestern.edu} \\
  \And
  Gregory J. Wagner \\
 Department of Mechanical Engineering\\
 Northwestern University\\
 Evanston, Illinois, 60201 \\
 \texttt{gregory.wagner@northwestern.edu} \\
 \And
 Wing Kam Liu \\
 Department of Mechanical Engineering\\
 Northwestern University\\
 Evanston, Illinois, 60201 \\
 \texttt{w-liu@northwestern.edu} \\
}

\begin{document}

\maketitle

\begin{abstract}
  A common trend in simulation-driven engineering applications is the ever-increasing size and complexity of the problem, where classical numerical methods typically suffer from significant computational time and huge memory cost. Methods based on artificial intelligence have been extensively investigated to accelerate partial differential equations (PDE) solvers using data-driven surrogates. However, most data-driven surrogates require an extremely large amount of training data. In this paper, we propose the Convolutional Hierarchical Deep Learning Neural Network-Tensor Decomposition (C-HiDeNN-TD) method, which can directly obtain surrogate models by solving large-scale space-time PDE without generating any offline training data. We compare the performance of the proposed method against classical numerical methods for extremely large-scale systems.
\end{abstract}

\section{Introduction}
Physics-based numerical simulation is the cornerstone in the rapidly evolving field of digital manufacturing. Despite the rapid development of computational hardware and many advances in high-performance computing (HPC) techniques, classical numerical algorithms still face several significant challenges for extremely large-scale problems that involve billions of degrees of freedom (DoF)\citep{li2024statistical, li2023convolution}, including long computational time, excessive memory consumption, and costly storage requirements. Deep learning-based numerical solvers, such as physics-informed neural networks (PINN), will also become expensive to solve long-time physical problems\citep{meng2020ppinn}.

Many data-driven methods have been proposed to address the bottleneck of classical numerical solvers \citep{huang2023introduction,li2020fourier,lu2021learning, saha2023deep, xie2022data}. Data-driven methods typically adopt the so-called offline-online approach. During the offline stage, high-fidelity data are generated from experiments or numerical simulations. Subsequently, various data-driven models can be used to approximate functional mapping from input to output spaces as surrogates \citep{li2020fourier, huang2023introduction, lu2021learning, park2024engineering}. Despite the success of these methods on various problems, it is still challenging to apply these data-driven methods to large-scale problems where offline training data generation can be very expensive\citep{li2024statistical}.


In order to circumvent the expensive offline data generation stage and improve the scalability of neural network structure for large-scale systems, we propose Convolutional Hierarchical Deep Learning Neural Networks-Tensor Decomposition (C-HiDeNN-TD) as an a priori method to directly obtain surrogates from solving large-scale PDE. In this paper, we will elucidate the key concepts behind C-HiDeNN-TD and showcase its advantages for large-scale simulations in terms of computational speed, memory efficiency, and storage requirements.

\section{Method}
\subsection{C-HiDeNN}
Thanks to the universal approximation theorem, multilayer perceptron (MLP) architectures have been successfully applied in deep learning-based solvers as global basis functions\citep{raissi2019physics}. However, MLP-based basis functions need special treatment to satisfy the boundary condition\citep{berrone2023enforcing}. In contrast, finite element (FE) basis functions can automatically satisfy the Dirichlet boundary condition due to the Kronecker-Delta property\citep{reddy1993introduction}. Moreover, numerical integration can be straightforward with Gaussian quadrature. In the previous work, we combined the concept of locally-supported FE basis functions and MLP to develop Convolutional-Hierarchical Deep Learning Neural Networks (C-HiDeNN) basis function. As shown in Eq. \ref{chidenn_sf}, $\bm{N_i}$ is the linear FE basis function at node $i$; $W$ is the kernel function that can be represented by a partially connected MLP\citep{lu2023convolution, park2023convolution}. This partially connected MLP is controlled by 3 hyperparameters: connectivity $s$, polynomial order $p$, and dilation parameter $a$. C-HiDeNN has been shown to be a flexible way to construct a high-order approximation with arbitrary convergence rates and automatic mesh adaptivity based on a linear FE mesh\citep{lu2023convolution}. More details of the C-HiDeNN formulation can be found in Appendix A. 
\begin{equation}
    u^{h}(\xi)=\sum_{i\in A^{e}}\bm{N}_{i}(\xi)\sum_{j\in A_{s}^{i}}\mathcal{W}_{s,a,p,j}^{i}\left(x^{h}(\xi)\right)\bm{u}_{j}=\sum_{k\in A_{s}^{e}}\bm{\widetilde{N}}_{k}(\xi)\bm{u}_{k}
\label{chidenn_sf}
\end{equation}

\subsection{C-HiDeNN Tensor decomposition}
Tensor decomposition (TD) is a popular method to decompose higher-order tensors into 1D vectors\citep{kolda2009tensor}. Likewise, a continuous multivariate function can be decomposed into univariate functions using TD\citep{lu2024extended}. In C-HiDeNN-TD, we decompose a multivariate function using the products of univariate functions approximated by the C-HiDeNN basis function. For an $n$-variate
function, the C-HiDeNN-TD approximation can be written as:
\begin{equation}
    u(\bm{p}) = u(p_1, p_2,...,p_I)\approx\sum_{m=1}^M\left[\widetilde{\boldsymbol{N}}_{p_1}(p_1)\boldsymbol{u}_{p_1}^{(m)}\right]\cdot\left[\widetilde{\boldsymbol{N}}_{p_2}(p_2)\boldsymbol{u}_{p_2}^{(m)}\right]\cdot...\cdot\left[\widetilde{\boldsymbol{N}}_{p_I}(p_I)\boldsymbol{u}_{p_I}^{(m)}\right]
\label{td_sum}
\end{equation}
where $m$ is defined as the decomposition rank; $M$ is the total rank number; $\bm{\widetilde{N}}_{p_i}$ is the 1D C-HiDeNN basis function for the $i$-th dimension; $\bm{{u}}^{(m)}_{p_i}$ is the solution vector of $i$-th dimension for rank $m$; $\cdot$ is the vector inner product;
Eq. \ref{td_sum} can also be written in a compact vector form:
\begin{equation}
    u(\bm{p})\approx\bm{[\widetilde{N}}_{p_1}(p_1)\bm{U}_{p_1}]\odot  \bm{[\widetilde{N}}_{p_2}(p_2)\bm{U}_{p_2}]\odot...\cdot\bm{[\widetilde{N}}_{p_n}(p_n)\bm{U}_{p_n}]
\end{equation}
where $\odot$ is the Hadamard product; $\bm{U}_{p_i}$ is the solution vector for all ranks: $\bm{U}_{p_i} = [\bm{{u}}^{(1)}_{p_i}, \bm{{u}}^{(2)}_{p_i}, ..., \bm{{u}}^{(M)}_{p_i}] $.

Data-driven TD has been extensively investigated as a data compression method for higher-order tensors\citep{kolda2009tensor, lu2019datadriven}. Unlike data-driven TD, a priori C-HiDeNN-TD can compute the surrogates (solution vectors $\bm{U}_{p_i}$) directly from PDE by circumventing the exorbitant offline high-resolution data generation process. Moreover, for high-dimensional problems, C-HiDeNN-TD will only have linear growth of the number of grid points compared to exponential growth in classical numerical algorithms. As a result, C-HiDeNN-TD can directly obtain surrogate models for exa-scale problems (number of grid points $> 10^{18}$) with significant speed-up, low memory cost, and efficient disk storage. More details about the C-HiDeNN-TD algorithm can be found in Appendix B.

\section{Results}
\subsection{Convergence study: Poisson equation}
In this section, we first study the convergence of C-HiDeNN-TD for the Poisson equation. The 2D Poisson equation is shown below.
\begin{equation}
    \left\{\begin{array}{ll}\Delta u(x,y)+b(x,y)=0 
 \quad \text{in} \quad \Omega_{(x,y)}\subset\mathbb{R}^2,\\u|_{\partial\Omega}=0.\end{array}\right.
\label{poisson}
\end{equation}
where $\Omega_{(x,y)}=[0,10]^2$ and homogeneous boundary condition is considered in the current example. We assume a concentrated load $b(x,y)=e^{\left[-10(x-5)^2-10(y-5)^2\right]}$. 
With C-HiDeNN-TD, the solution to Eq. \ref{poisson} can be approximated as:
\begin{equation}
    u^h(x,y)\approx\bm{[\widetilde{N}}_{x}(x)\bm{U}_{x}]\cdot\bm{[\widetilde{N}}_{y}(y)\bm{U}_{y}]
\end{equation}
To measure the accuracy of C-HiDeNN-TD, we define the energy norm error. 
\begin{equation}
    error=\frac{\|u^h-u^{ref}\|_E}{\|u^{ref}\|_E}=\frac{\sqrt{\int_\Omega\left|\nabla(u^h-u^{ref})\right|^2\mathrm{d}x}}{\sqrt{\int_\Omega\left|\nabla u^{ref}\right|^2\mathrm{d}x}}
\end{equation}
where the reference solution $u^{ref}$ is obtained using the linear finite element analysis with $8000 \times 8000$ elements.
\begin{figure}
  \centering
  \includegraphics[scale=0.48]{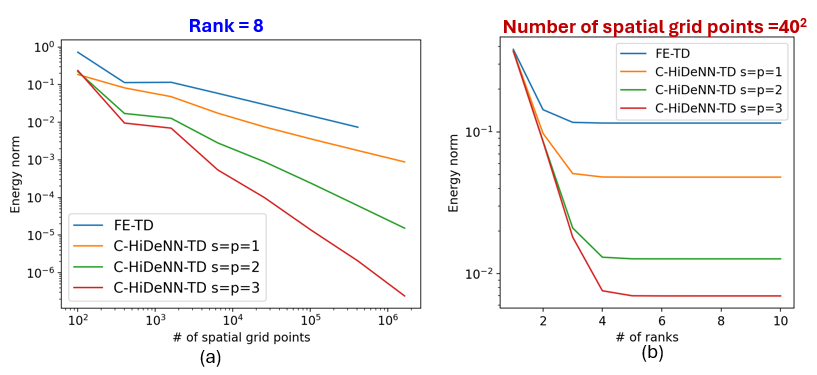}
  \caption{Convergence study of Poisson equation}
\end{figure}
The convergence plots are shown in Fig. 1, where number of spatial grid points is defined as the size of the original spatial mesh. In Fig. 1(a), it can be seen that the C-HiDeNN-TD has better accuracy compared to the FE-TD (FE basis function is used). Furthermore, C-HiDeNN-TD has a higher accuracy as connectivity ($s$) and polynomial order ($p$) of the neural network increase. This is because larger $s$ and $p$ will lead to higher-order smoothness of the basis function. As can be seen from Fig. 1(b), when the number of ranks increases, the energy norm error will decrease quickly for all cases until it reaches a plateau.

\subsection{Large scale transient diffusion analysis}

In this section, we investigate the performance of C-HiDeNN-TD for the large-scale transient diffusion problem. The governing equation is shown below, where the solution $u$ is a function of space and time.
\begin{equation}
    \left\{\begin{array}{ll}\rho c_p\dot u + k\Delta u+b(x,y,z)=0 
 \quad \text{in} \quad \Omega_{(x,y,z)}\otimes \Omega_{t},\\u|_{\partial\Omega}=0, \\ u|_{t=0}=0\end{array}\right.
\label{diffusion}
\end{equation}
where $\Omega_{(x,y,z)} = [0, 0.01]^3$; $\Omega_{t} = (0, 0.5]$; density $\rho = 1000$; heat capacity $c_p = 367.8$; heat conductivity $k = 5.836$; forcing term $b(x, y, z) = 2.55 \times  10^{11} \times  H(z - 9.5 \times 10^{-3})[e^{-2[\frac{(x-2\times10^{-3})^{2}+(y+2\times10^{-3})^{2}}{(5\times10^{-4})^{2}}]}+ e^{-2[\frac{(x-2\times10^{-3})^{2}+(y-2\times10^{-3})^{2}}{(5\times10^{-4})^{2}}]} + e^{-2[\frac{(x+2\times10^{-3})^{2}+(y-2\times10^{-3})^{2}}{(5\times10^{-4})^{2}}]}]$.

Unlike the classical semi-discretization approach in FE, where the spatial solution is approximated using FE basis function and the temporal problem is solved with time-stepping method, we directly use the C-HiDeNN-TD approximation for both space and time and solve the problem with a space-time approach.
\begin{equation}
    u^{h}(x,y,z,t)\approx\bm{[\widetilde{N}}_{x}(x)\bm{U}_{x}]\odot\bm{[\widetilde{N}}_{y}(y)\bm{U}_{y}]\odot\bm{[\widetilde{N}}_{z}(z)\bm{U}_{z}]\cdot\bm{[\widetilde{N}}_{t}(t)\bm{U}_{t}]
\end{equation}
The performance of C-HiDeNN-TD is compared with explicit finite difference method (FDM) in terms of computational time, GPU memory and disk storage. For fair comparison, both methods are coded in JAX and accelerated with just-in-time (JIT) compilation\citep{jax2018github}. Different discretizations have been used to study the scalability of the method. C-HiDeNN-TD uses the same discretization for all dimensions, whereas FDM uses the critical time increment for each case. All cases were tested on NVIDIA A6000 GPU with 48GB VRAM.

The computational time is compared in Fig. 2(b), where the $x$-axis is the total number of points for the spatial grids. For example, for the $100\times100\times100$ spatial grid, the total number of spatial grid points is $10^6$. As can be seen, the computational time of C-HiDeNN-TD is much less than FDM for large-scale problems. In terms of GPU memory usage, C-HiDeNN-TD requires much less memory as it decomposes the original 4D problem into a series of 1D problems. The largest space-time grids that C-HiDeNN-TD can resolve is $51,200^4 = 6.87\times 10^{18}$ (51,200 points along $x, y, z$ and $t$, respectively). Moreover, C-HiDeNN-TD also has a significant storage gain compared to FDM. Classical numerical methods like FDM store the spatial-temporal simulation data as a fourth-order tensor. However, C-HiDeNN-TD directly solves for 1D vectors, which are much cheaper to store. For the $51,200^4$ space-time grids, C-HiDeNN-TD only requires 15.6 MB to store the solution.
\begin{figure}
  \centering
  \includegraphics[scale=0.27]{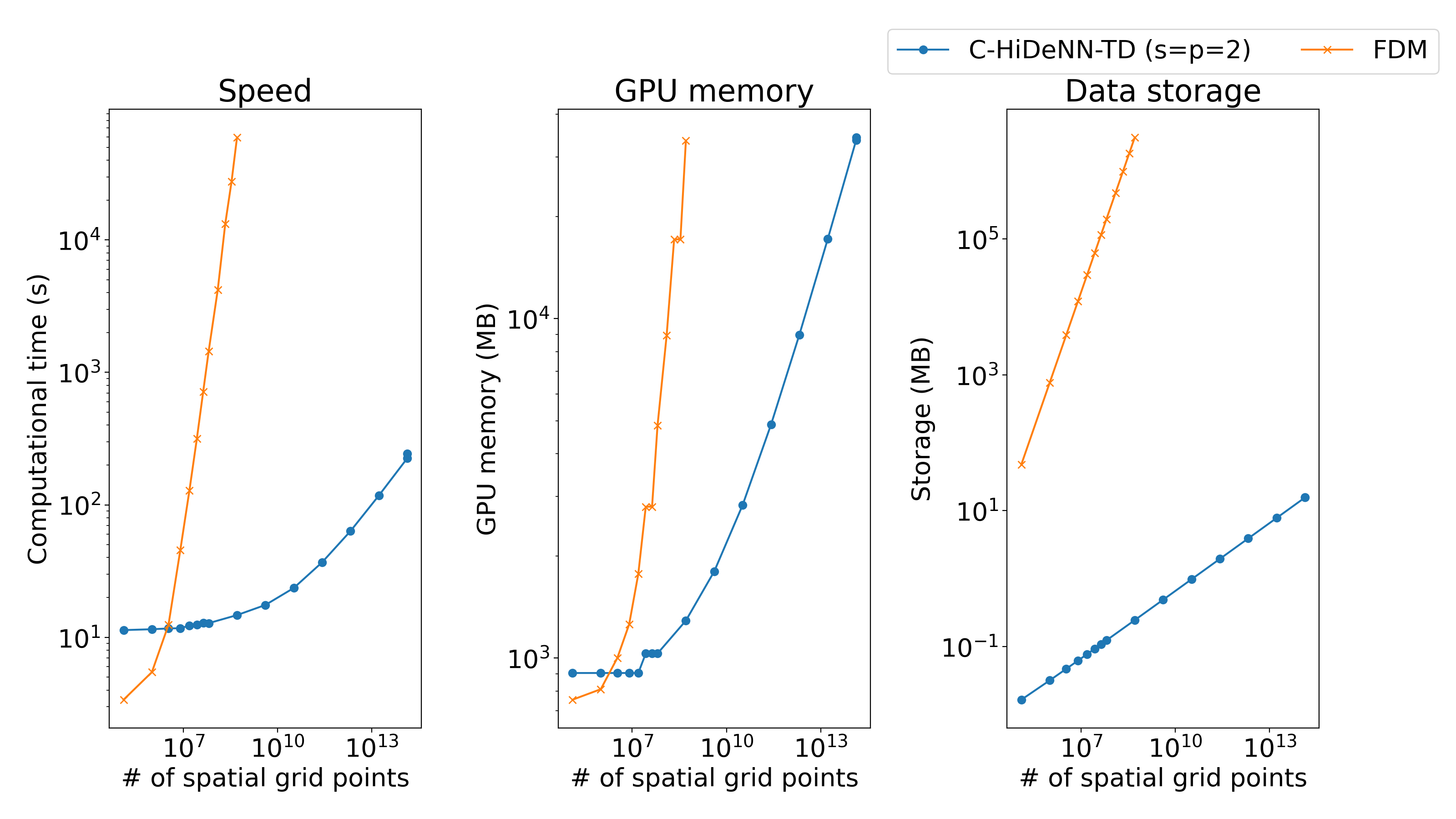}
  \caption{Comparison of C-HiDeNN-TD and finite difference method (FDM)}
\end{figure}



\section{Conclusion and future work}
In this paper, we proposed a new type of neural network-based solver, C-HiDeNN-TD, to tackle the challenge of the ever-growing size of computational problems in real engineering applications. C-HiDeNN-TD can obtain surrogates without the expensive high-fidelity simulation data generation process. Moreover, C-HiDeNN-TD can degenerate high-dimensional problems into 1D subproblems. We have demonstrated the performance of C-HiDeNN-TD for steady-state/transient diffusion equations. C-HiDeNN-TD can solve for surrogate models for large-scale problems without generating expensive data. The current method can be easily extended to a more general space-time-parameter (S-T-P) formulation , where the parameters of PDE can also be included as extra-coordinates\citep{lu2024extended}. Moreover, C-HiDeNN-TD will be applied to solve nonlinear PDE such as Navier-Stokes equation.

\newpage
\bibliographystyle{unsrtnat}
\bibliography{ref} 
\section*{Appendix}
\subsection*{A. C-HiDeNN theory} The
C-HiDeNN basis function exploits the merits of the finite element basis function and the flexibility of multilayer perceptron (MLP). The first first part of C-HiDeNN is the Lagrange polynomial in finite element methods, which can exactly satisfy the Dirichlet boundary condition. As shown in the previous work\citep{zhang2021hierarchical}, 1st-order Lagrange polynomial is equivalent to a partially connected 2-layer MLP. To further enhance the smoothness, we add an additional third layer with nonlinear activation with $p$ reproducing polynomial order. The full mathematical form of C-HiDeNN basis function can be written as:
\begin{equation}
    u^{h,e}(\xi)=\sum_{i\in A^{e}}\bm{N}_{i}(\xi)\sum_{j\in A_{s}^{i}}\mathcal{W}_{s,a,p,j}^{i}\left(x^{h, e}(\xi)\right)\bm{u}_{j}=\sum_{k\in A_{s}^{e}}\bm{\widetilde{N}}_{k}(\xi)\bm{u}_{k}
\label{chidenn}
\end{equation}
where $\bm{N}_i (\xi)$ is polynomial interpolant at node $i$, $W^i_{s,a,p,j} (\xi)$ is the convolution patch function at node $j$ defined on the nodal patch domain $A_s^i$ with three hyper-parameters ($s$: patch size, $a$: dilation parameter, $p$: polynomial order), and $x^{h,e} (\xi)$ is the FEM mapping from the parent coordinate $\xi$ to the physical coordinate $x$ in element $e$. Note that the input of the neural network is changed from the physical coordinate $x$ to the parent coordinate $\xi$ in Fig. 3(a) to facilitate the numerical integration of the weak form of PDE. The convolution patch functions are precomputed from the meshfree theory and used as weights between the second and third hidden layers in Figure 3(b). More details on how the weights are calculated can be found in \citep{park2023convolution}.

\begin{figure}[h]
  \centering
  \includegraphics[width=0.5\linewidth]{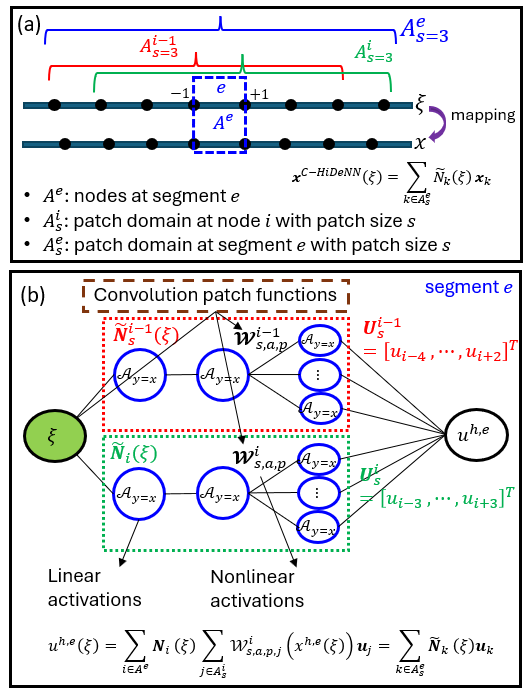}
  \label{a1}
  \caption{C-HiDeNN structure. (a) 1D C-HiDeNN basis function is constructed by aggregating the information from all the segments patch domain into the current segment (b) Neural network structure of C-HiDeNN basis function: the first 2 hidden layers are mapping the input to piece-wise continuous function $N$, which satisfies the Kronecker delta property; the 3rd layer is the nonlinear part that acts as the convolution patch function $W$, which satisfies the reproducing condition; the output field variable $u^{h,e}$ has higher-order smoothness and automatically satisfy the Dirichlet boundary condition thanks to the Kronecker delta property.}
\end{figure}

\subsection*{B. A priori C-HiDeNN-TD}
In this section, the algorithm of C-HiDeNN-TD is explained for the 2D Poisson equation. C-HiDeNN-TD solves the weak form of the Poisson equation using the Galerkin formulation. The weak form of 2D Poisson equation is:
\begin{equation}
    \int_{\Omega}\bm\nabla\delta u\cdot \bm\nabla u-\int_{\partial\Omega}\delta u\cdot\left(\bm\nabla u\cdot \bm{n}\right)-\int_{\Omega}\delta ub=0
\label{weak}
\end{equation}
where $u$ is the trial function, $\delta u$ is the test function. In Galerkin formulation, we use the same C-HiDeNN-TD basis function for the trial and test functions. Note that we have 2 test functions for $x$ and $y$ dimension, respectively.
\begin{equation}
    u(x,y)\approx\bm{[\widetilde{N}}_{x}(x)\bm{U}_{x}]\cdot\bm{[\widetilde{N}}_{y}(y)\bm{U}_{y}]
    \label{trial}
\end{equation}

\begin{subequations}
\begin{equation}
    \delta u_x(x,y)\approx\bm{[\widetilde{N}}_{x}(x)\bm{\delta U}_{x}]\cdot\bm{[\widetilde{N}}_{y}(y)\bm{U}_{y}]
    \label{testx}
\end{equation}

\begin{equation}
    \delta u_y(x,y)\approx\bm{[\widetilde{N}}_{x}(x)\bm{ U}_{x}]\cdot\bm{[\widetilde{N}}_{y}(y)\bm{\delta U}_{y}]
    \label{testy}
\end{equation}
\label{test}
\end{subequations}
Plugging Eq. \ref{trial}-\ref{test} into Eq. \ref{weak} and simplifying the equations, The original 2D problem is decomposed into two 1D subproblems as shown below.
\begin{subequations}
\begin{equation}
    \bm{{A}_{x}{U}_{x}={Q}_{x}} 
\label{td1}
\end{equation}

\begin{equation}
\bm{{A}_{y}{U}_{y}={Q}_{y}}
\label{td2}
\end{equation}
\end{subequations}
where $\bm{A_i} \subset \mathbb{R}^{ML_i \times ML_i}$ is the coefficient matrix for dimension $i$ ($M$ refers to number of ranks; $L_i$ is the number of nodes in the $i$-th dimension); $\bm{U_i} \subset \mathbb{R}^{ML_i}$ is the solution vector. Eq. \ref{td1} and Eq. \ref{td2} are solved iteratively until the variation of $\bm{{U}_{x}}$ and $\bm{{U}_{x}}$ is within the tolerence range.

\end{document}